    \documentclass[prd,aps,preprint,tightenlines,showpacs,nofootinbib,superscriptaddress]{revtex4-1}
\usepackage{mathrsfs}
\usepackage{amsfonts}
\usepackage{amsmath}
\usepackage{amssymb}
\usepackage{array}
\usepackage{verbatim}
\usepackage{bm}
\usepackage{epsfig}
\usepackage{graphicx,color}
\usepackage{relsize}
\usepackage{lineno}
\usepackage{float}
\usepackage{multirow}
\RequirePackage{xspace}

\def\lsim{\raise0.3ex\hbox{$<$\kern-0.75em\raise-1.1ex\hbox{$\sim$}}}

\def\gsim{\raise0.3ex\hbox{$>$\kern-0.75em\raise-1.1ex\hbox{$\sim$}}}

\newcommand{\be}{\begin{equation}}

\newcommand{\ee}{\end{equation}}

\def\beq{\begin{equation}}

\def\eeq{\end{equation}}

\def\beqa{\begin{eqnarray}}

\def\eeqa{\end{eqnarray}}

\newcommand{\ba}{\begin{eqnarray}}

\newcommand{\ea}{\end{eqnarray}}

\def\gappeq{\mathrel{\rlap {\raise.5ex\hbox{$>$}}

{\lower.5ex\hbox{$\sim$}}}}

\def\lappeq{\mathrel{\rlap{\raise.5ex\hbox{$<$}}

{\lower.5ex\hbox{$\sim$}}}}

\def\Toprel#1\over#2{\mathrel{\mathop{#2}\limits^{#1}}}

\begin{document}

\title{Probing the pion gluon distribution at small - $x$
in photon-induced interactions at LHC}

\author{Victor P. {\sc Gon\c{c}alves}}
\email{barros@ufpel.edu.br}
\affiliation{Institute of Physics and Mathematics, Federal University of Pelotas (UFPel), \\
  Postal Code 354,  96010-900, Pelotas, RS, Brazil}

\author{Juciene T. de {\sc Souza}}
\email{juciteixeiraprof@gmail.com}
\affiliation{Institute of Physics and Mathematics, Federal University of Pelotas (UFPel), \\
  Postal Code 354,  96010-900, Pelotas, RS, Brazil}

\author{Diego  {\sc Spiering}}
\email{diego.spiering@gmail.com}
\affiliation{Instituto de F\'{\i}sica, Universidade de S\~{a}o Paulo,
C.P. 66318,  05315-970 S\~{a}o Paulo, SP, Brazil}

\begin{abstract}
In this paper, we propose the analysis of the heavy quark photoproduction associated with a leading neutron in hadronic collisions at the LHC as an alternative to probe the pion gluon distribution in a kinematical range not covered by previous experiments.  We perform an exploratory study of the charm and bottom photoproduction associated with a leading neutron in proton-proton ($pp$) and proton-lead ($pPb$) collisions, and present calculations for the  rapidity distributions and cross-sections.
Our predictions indicate that experimental studies of these processes are feasible, and that a future measurement of this final state will be useful to
constrain the pion gluon distribution at small values of the Bjorken $x$ variable and to improve our understanding of the pion structure.
\end{abstract}



\maketitle

\vspace{1cm}

\section{Introduction}
Over the last decades, the study of particle production by photon-photon and photon-hadron interactions in hadronic collisions at the LHC became a reality and is currently one of the more promising ways to investigate the proton and nucleus structure as well to probe for signals of Beyond the Standard Model Physics \cite{upc}.
During this period, the four experimental LHC collaborations have released data for exclusive processes (e.g. the exclusive vector-meson photoproduction), where both incident particles remain intact,  mainly motivated by the possibility of improving the description of QCD dynamics at high energies \cite{klein,gluon,Frankfurt:2001db}. More recently, the first data for inclusive processes, where one of the incident hadrons is broken up, have also been released~\cite{ATLAS:2024mvt,CMS:2025jjx,Nese:2025ohz},  providing additional constraints on the description of the hadronic structure as indicated e.g. in Refs. \cite{Hofmann:1991xza,Baron:1993nk,Greiner:1994db,gluon,Klein:2000dk,Klein:2002wm,Goncalves:2003is,Goncalves:2004dn,Goncalves:2009ey,Goncalves:2013oga,Goncalves:2015cik,Kotko:2017oxg,Goncalves:2017zdx,Guzey:2019kik,Guzey:2018dlm,Eskola:2024fhf,Goncalves:2019owz,Gimeno-Estivill:2025rbw,Goncalves:2025wwt,Cacciari:2025tgr}. In particular, the recent studies performed in Refs.~\cite{Gimeno-Estivill:2025rbw,Goncalves:2025wwt,Cacciari:2025tgr}   have demonstrated that the analysis of inclusive heavy quark photoproduction in ultraperipheral collisions (UPCs) is an important probe of the gluon distribution of the proton or nucleus target at small values of the Bjorken - $x$ variable.

\begin{figure}[t]
	\centering
\includegraphics[width=0.8\textwidth]{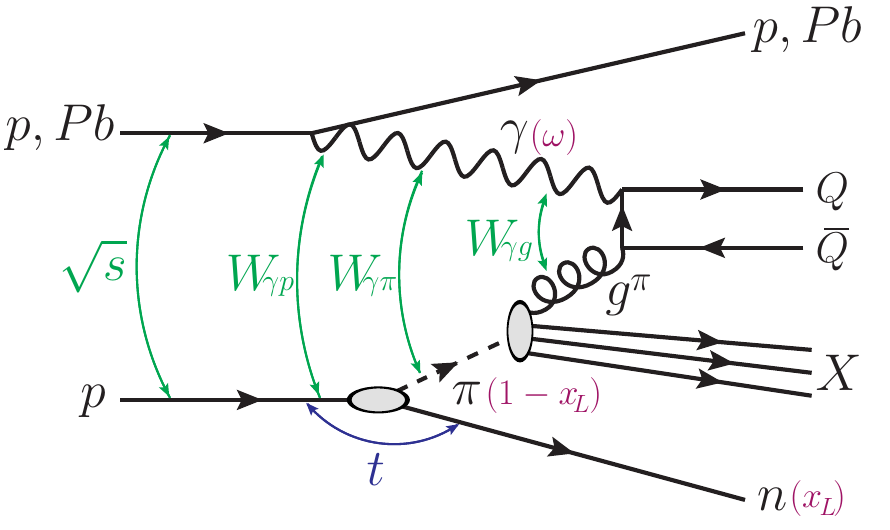} 
\caption{Heavy quark photoproduction by photon - pion interactions at $pp$ and $pPb$  collisions. }
\label{fig:diagram}
\end{figure}

Our goal in this paper is to investigate, for the first time, if the study of inclusive heavy quark photoproduction in UPCs can also be used to probe the pion structure in a kinematical range complementary to that accessed in pion - nucleus scattering at CERN and Fermilab \cite{NA3:1983ejh,NA10:1985ibr,E615:1989bda}, and  leading neutron electroproduction at HERA \cite{ZEUS:2002gig,H1:2010hym}. 
In our analysis, we will consider the Sullivan process~\cite{Sullivan:1971kd}, in which the photon emitted by one of the incoming hadrons scatters off the virtual pion cloud of the hadron target, effectively probing the pion structure. In particular, the heavy quark photoproduction cross-section will be determined by the gluonic content of pion (See Fig. \ref{fig:diagram}). Moreover, the final state will be characterized by the presence of a rapidity gap and a leading neutron, which can be tagged using a Zero Degree Calorimeter (ZDC). In our analysis, we will consider the charm and bottom photoproduction in $pp$ and $pPb$ collisions at the LHC, and will estimate the corresponding rapidity and total cross-sections assuming different parameterizations for the pion gluon distribution  available in the literature \cite{Gluck:1991ey,JeffersonLabAngularMomentumJAM:2022aix,Novikov:2020snp}. Finally, we will show that the ratio between the charm and bottom  rapidity distributions is a sensitive probe of the pion gluon distribution, which can be used to constrain the behavior of this distribution. 
As we will demonstrate below, the analysis of inclusive process with a leading neutron in UPCs provides an important probe of pion structure, complementary to that obtained by the study of exclusive processes with a leading neutron, originally proposed in Refs.~\cite{Goncalves:2016uhj,Carvalho:2017vtw}.

This paper is organized as follows. In the following section, we present a brief review of the  formalism needed to describe the heavy quark photoproduction associated with a leading neutron in UPCs. In particular, we will discuss the description of the  photon and pion fluxes, as well as the expression for the heavy quark photoproduction cross-section.  In section \ref{sec:results}, the $x$ - dependence of the pion gluon distribution predicted by the distinct parameterizations used in our calculations will be presented, and we will present our results for the energy dependence of the  cross-sections for the charm and bottom production in photon - pion interactions. Moreover, predictions for the rapidity distributions and total cross-sections associated with the heavy quark photoproduction in $pp$ and $pPb$ collisions   will be shown. In the last section \ref{sec:sum}, we summarize our main predictions and conclusions.

\section{Formalism}
In this section, we will present a brief review of the formalism needed to describe the heavy quark photoproduction associated with a leading neutron (HQ+LN) in ultraperipheral collisions. { In this exploratory study, we will focus on the direct contribution,  in which the photon interacts as a point-like particle, and postpone the inclusion of the  resolved processes, where it fluctuates into a hadronic state, for a future publication (For a detailed discussion see, e.g., Ref.~\cite{Klasen:2002xb}).} The { direct photoproduction} process is represented in Fig. \ref{fig:diagram} for $pp$ and $pPb$ collisions, where we also present the main kinematical variables. The basic idea is that in a hadronic collision with center-of-mass energy $\sqrt{s}$, one of the incident protons (or nucleus) can be considered as a source of photons of energy $\omega$, which will interact with the proton target with a squared center-of-mass energy $W^2_{\gamma p} = 2 \omega \sqrt{s}$ \cite{upc}. Assuming the Sullivan process~\cite{Sullivan:1971kd}, the photon scatters off the pion cloud of the proton target, with the squared center-of-energy for the photon-pion interaction being given by ${W}_{\gamma \pi}^2 = (1-x_L) \, W_{\gamma p}^2$, where $x_L$ is the proton momentum fraction carried by the 
neutron. Moreover, $t$ is the square of the four-momentum of the exchanged pion, which can be expressed in terms  of the measured quantities $x_L$ and 
transverse momentum $p_T$ of the neutron, as follows:
\beq
t \simeq-\frac{p_T^2}{x_L}-\frac{(1-x_L)(m_n^2-m_p^2 x_L)}{x_L} \,\,,
\label{virtuality}
\eeq
where $m_p$ $(m_n)$ is the proton (neutron) mass. In the case of the heavy quark photoproduction, the dominant partonic channel is the photon - gluon interaction, $\gamma g \rightarrow Q \bar{Q}$, with the gluon in the pion $g^\pi$ carrying a momentum fraction $x$, which implies  ${W}_{\gamma g}^2 = x \, W_{\gamma \pi}^2$. As a consequence, the corresponding cross-section will be sensitive to the pion gluon distribution $xg^{\pi}(x,\mu^2)$ at a given value of the Bjorken $x$ variable and hard scale $\mu \propto m_Q$, where $m_Q$ is the heavy quark mass.
The final state for the HQ + LN  process in UPCs  will be characterized by a rapidity gap, associated with the photon exchange, the heavy quark pair and a leading neutron, which carries a large fraction of the proton energy. In principle, such a process can be separated from the dominant heavy quark production by photon-proton interactions by  tagging the forward neutron in the final state.

Assuming the equivalent photon approximation \cite{epa}, the rapidity distribution for the HQ+LN process in an ultraperipheral $h_1h_2$ collision will be given by
\begin{eqnarray}
  \frac{d\sigma \,[h_1+h_2 \rightarrow h_i+Q\bar{Q}+X+n]}{dY} = 
  \left[ \omega \frac{dN}{d\omega}\bigg|_{h_1}\,\sigma_{\gamma h_2 \rightarrow Q\bar{Q} + X + n}\left(\omega \right) \right]_{\omega_L} +
  \left[ \omega \frac{dN}{d\omega}\bigg|_{h_2}\,\sigma_{\gamma h_1 \rightarrow Q\bar{Q} + X + n}\left(\omega \right) \right]_{\omega_R}\,, \nonumber \\
  \label{eq:rapdis}
  \end{eqnarray}
where $Y$ is the rapidity of the heavy quark pair, $h_i$ represents the hadron that have emitted the photon and remains intact in the final state, $n$ is the leading neutron and $X$ is the hadronic state generated by the pion break up. Moreover, ${dN}/{d\omega}|_{h_j}$ is the equivalent photon flux associated with the hadron $h_j$ and $\sigma_{\gamma h \rightarrow Q\bar{Q} + X + n}$ is the cross-section for the heavy quark photoproduction associated with a leading neutron. Finally, $\omega_L \propto \exp{(+Y)}$ and $\omega_R \propto \exp{(-Y)}$ denote photons associated with the hadrons  $h_1$ and $h_2$, respectively.

In our analysis, we will focus on $pp$ and $pPb$ collisions and assume that the equivalent photon fluxes associated with the nucleus and  proton are given by \cite{upc}
\begin{eqnarray}
  \frac{dN_{\gamma/A}\,(\omega)}{d\omega}= \frac{2\,Z^2\alpha_{em}}{\pi\,\omega}\, \left[\bar{\eta}\,K_0\,(\bar{\eta})\, 
  K_1\,(\bar{\eta})+ \frac{\bar{\eta}^2}{2}\big[K_0^2\,(\bar{\eta}) - K_1^2\,(\bar{\eta})\big] \right]\,,\\
\end{eqnarray}
and 
\begin{eqnarray}
  \frac{dN_{\gamma/p}\,(\omega)}{d\omega}= \frac{\alpha_{em}}{2\pi\,\omega}\, \left[1 + \left( 1-\frac{2\omega}{\sqrt{s}}\right)^2\right]
  \left( \ln\Omega - \frac{11}{6} + \frac{3}{\Omega} - \frac{2}{3\Omega^2} + \frac{1}{3\Omega^3} \right)\,
\end{eqnarray}
where $\omega=m_Qe^Y$, $\bar{\eta}=\omega\,(R_{A} + R_{p})/\gamma_L$, 
$\Omega = 1 + [(0.71\text{ GeV}^2)/Q^2_{min}]$ and 
$Q^2_{min}=\omega^2/[\gamma_L^2(1-2\omega/\sqrt{s})]$, with $K_0$ and $K_1$ being  the 
modified Bessel functions and $\gamma_L$ the Lorentz factor. For $pPb$ collisions, the cross-section is dominated by $\gamma p$ interactions due to the factor $Z^2$ in the nuclear photon flux. 

In order to describe the cross-section $\sigma_{\gamma h \rightarrow Q\bar{Q} + X + n}$ we will assume the Sullivan process~\cite{Sullivan:1971kd}, which implies that it can be expressed as follows (See e.g. Ref.~\cite{nos1}) 
\begin{eqnarray}
  \sigma_{\gamma p \rightarrow Q\bar{Q} + X + n} (W^2_{\gamma p}) = {\cal{K}} \cdot \int dx_L dt \, f_{\pi/p} (x_L,t) \cdot 
  \sigma_{\gamma \pi \rightarrow Q\bar{Q} + X}(W^2_{\gamma\pi})\,,
\end{eqnarray}
where ${\cal{K}}$ represents the absorption factor  associated to soft rescatterings between the produced and spectator particles\footnote{For a detailed discussion about the absorptive effects see Ref. \cite{Carvalho:2020myz}.} and $f_{\pi/p} (x_L,t)$ is the pion flux. The main assumption here is that  the splitting  $p \rightarrow \pi^+ n$, the photon -- pion
interaction and the absorptive effects can be factorized. 
The general form of the pion flux is given by
\beq
f_{\pi/p} (x_L,t)  = \frac{1}{4 \pi} \frac{2 g_{p \pi p}^2}{4  \pi} \frac{-t}{(t-m_{\pi}^2)^2} (1-x_L)^{1-2 \alpha(t)}  
[F(x_L,t)]^2
\label{genflux}
\eeq 
where $g_{p \pi p}^2/(4 \pi) = 14.5$ is the $p \pi p $ coupling constant, $m_{\pi}$ is the pion mass, $\alpha(t)$ is the pion intercept and the form factor $F(x_L,t)$  accounts for the finite size of the nucleon and pion. Following Refs. \cite{kope,nos1}, we will assume an exponential form factor given by
\beq
F(x_L,t) =  \exp [ b (t - m_{\pi}^2) ] \,\,\,\, , \,\,\,\, \alpha(t) = t
\label{form3}
\eeq
 with $b = 0.3$ GeV$^{-2}$. As demonstrated in Refs.  \cite{kope,nos1}, such choice allow us to describe the current leading neutron electroproduction HERA data~\cite{ZEUS:2002gig,H1:2010hym}.  It is important to emphasize that assuming the validity of the factorization hypothesis and the universality of the fragmentation process, the data of leading neutron production in $pp$ collisions can be used to constrain $f_{\pi/p}$, reducing the dependence of our predictions on this assumption. Finally, in this exploratory study, the heavy quark photoproduction in photon-pion interactions will be estimated at leading order, which implies that the associated cross-section will be given by~\cite{Gluck:1978bf}
\begin{eqnarray}
  \sigma^{\gamma \pi \rightarrow Q\bar{Q}X}(W_{\gamma \pi}^2) = 
  \int_{x_{min}}^1 dx \,\, \sigma^{\gamma g\rightarrow Q\bar{Q}}(W^2_{\gamma g})\,\,g^{\pi}(x, \mu^2)\,,
\end{eqnarray}
with $x_{min} = {4m_Q^2}/{W_{\gamma \pi}^2}$ and
\begin{eqnarray}
  \sigma^{\gamma g\rightarrow Q\bar{Q}}(W^2_{\gamma g}) = \frac{2\pi\alpha_{em}\alpha_s(\mu^2)e_Q^2}{{W}_{\gamma g}^2}
  \left[
    \left( 1 + \beta -\frac{\beta^2}{2} \right) \ln\left(\frac{1+\nu}{1-\nu}\right) -(1+\beta)\nu
  \right]\,,
  \label{gamma_gluon}
\end{eqnarray}
where $\nu = \sqrt{1-\beta}$  and $\beta = {4m_Q^2}/{W_{\gamma g}^2}$. In our calculations, we will assume $\mu=2\,m_Q$, with $m_c = 1.5$ GeV and $m_b = 4.5$ GeV, and ${\cal{K}} = 0.8$, which allow us to describe the HERA data for the leading neutron electroproduction \cite{nos1}.

\begin{figure}[t]
	\centering
\includegraphics[width=1.0\textwidth]{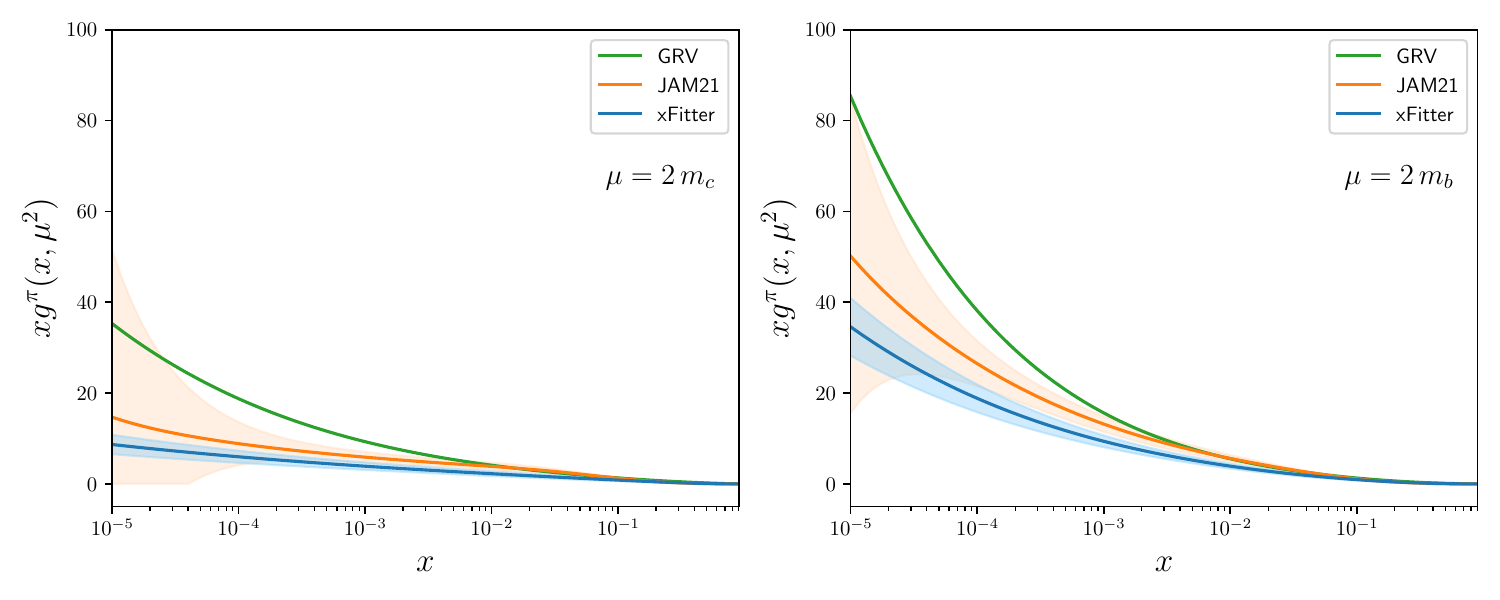} 
\caption{Comparison between the pion gluon distributions predicted by the GRV, JAM21 and xFitter parameterizations~\cite{Gluck:1991ey,JeffersonLabAngularMomentumJAM:2022aix,Novikov:2020snp}. Results for two distinct values of the hard scale $\mu$.}
\label{fig:gluon}
\end{figure}

\section{Results}
\label{sec:results}
In what follows, we will present our results for the rapidity distributions and cross-sections for the charm and bottom photoproduction associated with a leading neutron in $pp$ collisions at $\sqrt{s} = 13$ TeV and $pPb$ collisions at $\sqrt{s} = 8.1$ TeV. Such predictions will be derived assuming three different parameterizations for the pion gluon distribution, which are available in the LHAPDF library\footnote{https://www.lhapdf.org/}. In particular, we will consider the GRV~\cite{Gluck:1991ey}, JAM21~\cite{JeffersonLabAngularMomentumJAM:2022aix} and xFitter~\cite{Novikov:2020snp} parameterizations, which are based on different assumptions for the initial conditions of the DGLAP evolution and consider distinct data sets to constrain the free parameters. In Fig. \ref{fig:gluon} we present a comparison between the predictions of these parameterizations for the gluon distribution, derived considering two distinct values of the hard scale $\mu$. In addition to the central value, the uncertainty band on the JAM21 and xFitter gluon parameterizations is also presented. Such uncertainty is not available in the GRV case. We have that all parameterizations predict the increase of the distribution at small - $x$, but the increasing is dependent on the parameterization. Moreover, the uncertainty band increases for smaller values of $x$, which is directly associated with the fact that the  current data for  Drell - Yan and leading neutron DIS processes are not able to constrain the pion gluon distribution in the kinematical range. 

\begin{figure}[b]
	\centering
\includegraphics[width=1.0\textwidth]{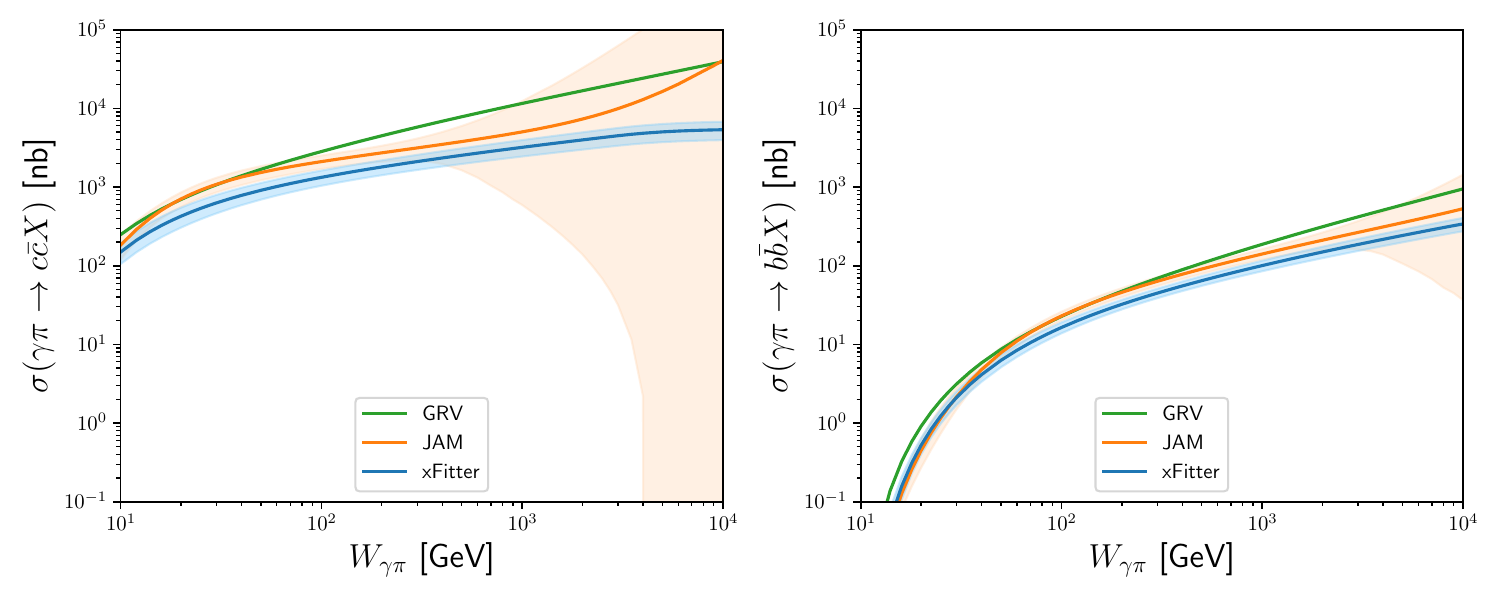} 
\caption{Energy dependence of  the total cross - section for  the charm (left panel) and bottom (right panel) pair production  in photon - pion interactions.  }
\label{fig:photopion}
\end{figure}

The impact of the distinct parameterizations for the pion gluon distributions on the energy dependence of the cross-section for the heavy quark photoproduction in photon - pion interactions is presented in Fig. \ref{fig:photopion}. Results for the charm (bottom) pair production are presented in the left (right) panel. The cross-sections increase with the energy, since  smaller values of $x$ are probed at larger $W$, and decrease for heavier quarks. Moreover, the predictions are sensitive to the parameterization assumed as input in the calculations. 

Let's now consider ultraperipheral collisions at the LHC. As discussed in the Introduction, in these collisions, the incoming hadrons can be considered as sources of photons of energy $\omega$, whose distribution is described by the equivalent photon flux. 
The maximum photon energy can  be derived considering that the maximum possible momentum in the longitudinal direction is modified by the Lorentz factor, $\gamma_L$, due to the Lorentz contraction of the hadrons in that direction \cite{upc}. It implies $\omega_{\mbox{max}} \approx \gamma_L/R_h$ and, consequently, $W_{\gamma p}^{\mbox{max}} = \sqrt{2\,\omega_{\mbox{max}}\, \sqrt{s}}$. Therefore,  for $pp/pPb$ collisions at $\sqrt{s} = 13 / 8.1 $ TeV, we have that the maximum photon-proton center-of- mass energy, $W_{\gamma p}^{\mbox{max}}$, will be of the order of  $8.2 /1.4$ TeV \cite{upc}, i.e.,  LHC  probes a range of  photon-proton center-of-mass energies unexplored by HERA. Such a conclusion is also valid for photon-pion interactions, which implies  that the study of heavy quark photoproduction associated with a leading neutron in UPCs allow us to investigate the pion structure in the high energy limit.

\begin{figure}[t]
	\centering
\includegraphics[width=1.0\textwidth]{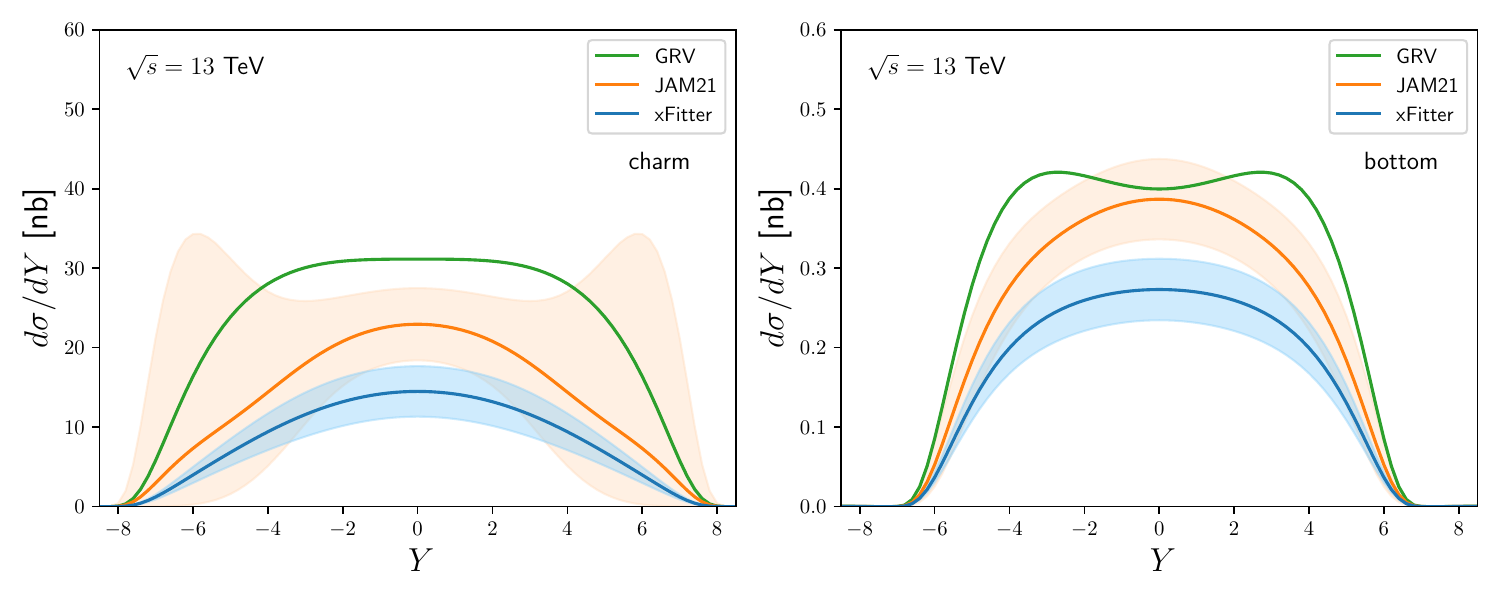}
\caption{Rapidity distributions for the charm (left panel) and bottom (right panel) pair photoproduction associated with a leading neutron in $pp$ collisions at $\sqrt{s} = 13$ TeV. Results derived assuming different parameterizations for the pion gluon distribution.  }
\label{fig:dsdypp}
\end{figure}

In Fig. \ref{fig:dsdypp} we present our predictions for the rapidity distributions  for the charm and bottom photoproduction associated with a leading neutron in $pp$ collisions at the LHC, derived assuming different parameterizations for the pion gluon distribution. The results are obtained by calculating Eq. (\ref{eq:rapdis}). 
The first term in Eq. (\ref{eq:rapdis}) is determined by the photon flux for a photon with  energy $\omega \propto e^{Y}$ and the   heavy quark photoproduction cross-section for a given photon-proton center-of-mass energy $W_{\gamma p}$. While $\sigma_{\gamma p \rightarrow Q\bar{Q} + X + n}$ increases with $W_{\gamma p}$, the  photon flux strongly decreases when the photon energy is of the order of $\omega_{\mbox{max}} \approx \gamma_L/R_p$, becoming almost zero for larger photon energies. As a consequence,  this contribution  increases with the rapidity up to a maximum and  becomes zero at very large $Y$. On the other hand, the second term in Eq. (\ref{eq:rapdis}) increases for negative values of rapidity, since in this case $\omega \propto e^{-Y}$. For $pp$ collisions, the contributions of both terms in Eq. (\ref{eq:rapdis}) are identical and symmetric in rapidity.   Moreover, we have that the increasing with rapidity is determined by the energy dependence of the heavy quark photoproduction cross-section, being dependent on the pion gluon distribution considered. Therefore, the difference between the predictions presented in Fig.\ref{fig:dsdypp} is larger for larger values of $|Y|$. However, the results are also distinct at midrapidities ($Y \approx 0$), with the central predictions  for the charm production associated with the JAM21 and xFitter parameterizations differing by a factor $\approx 2$.  A similar difference is also observed in the case of bottom production (right panel).
{ It is important to emphasize that it should be possible to identify
the pion-going direction by tagging the forward neutron, which will allow us to use this information to separate the contribution of each term in Eq. (\ref{eq:rapdis}), allowing a more precise determination of the pion gluon distribution since the ambiguity about what of the incident protons was the photon emitter is eliminated. Such aspect will be explored in a future publication. }

\begin{figure}[t]
	\centering
\includegraphics[width=1.0\textwidth]{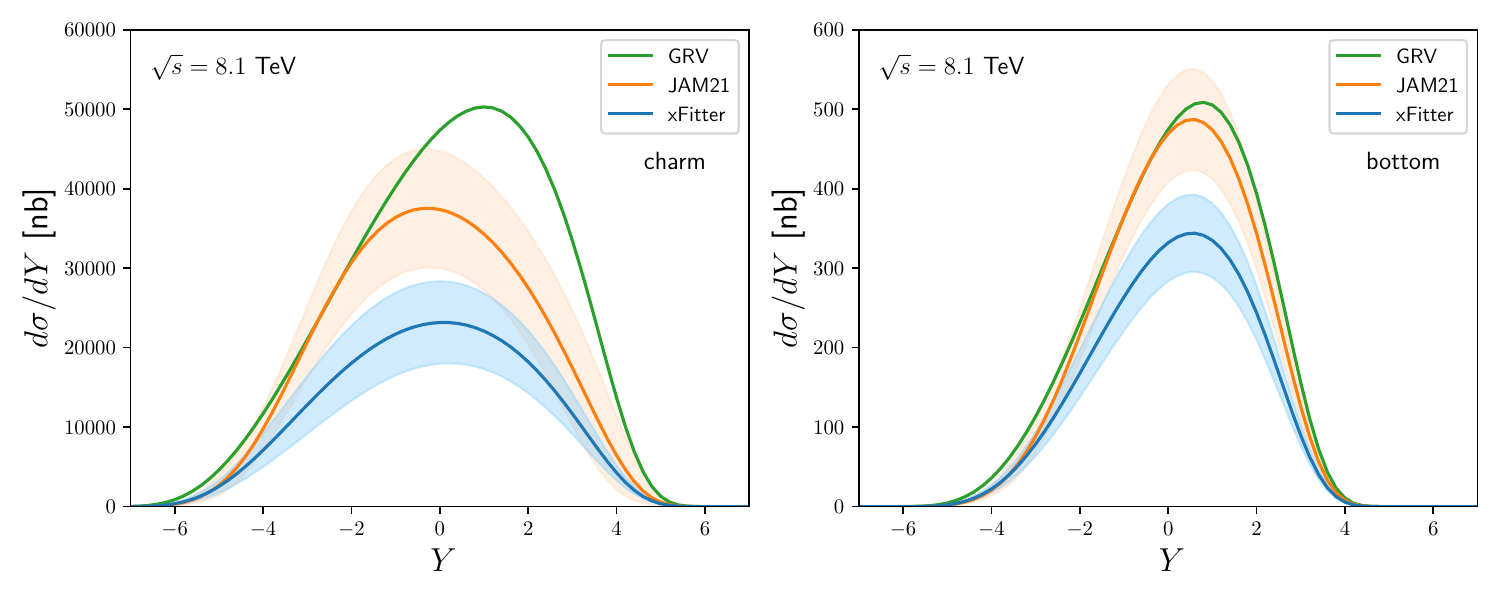}
\caption{Rapidity distributions for the charm (left panel) and bottom (right panel) pair photoproduction associated with a leading neutron in $pPb$ collisions at $\sqrt{s} = 8.1$ TeV. Results derived assuming different parameterizations for the pion gluon distribution.   }
\label{fig:dsdypA}
\end{figure}

In Fig. \ref{fig:dsdypA} we present our results for $pPb$ collisions. In this case, the rapidity distribution receives contributions of photon-proton and photon-nucleus interactions. However, due to the $Z^2$ factor present in the nuclear photon flux, the distribution is dominated by the photon-proton contribution. As a consequence,  the associated rapidity distribution is asymmetric and larger in magnitude in comparison with the $pp$ case. We have that the difference between the predictions increases with the rapidity, with the position of the maximum of distribution being dependent on the PDF parameterization.

\begin{table}[t]
    \vspace{-0.5cm}
    \centering
    \begin{tabular}{|c|c|c|l|l|} 
        \hline
        \hline
        Final state & Rapidity range & PDF Parameterization & $\sigma(p\text{Pb})$ [$\mu$b] & $\sigma(pp)$ [nb]\\ 
        \hline
        \hline
        \hline
        $c\bar{c}$ & [-7, 7] & GRV & $3.02\times 10^{2}$ & $3.47\times 10^{2}$\\ 
                \hline
        $c\bar{c}$ & [-7, 7] & JAM21 & $2.24\times 10^{2}$ & $2.14\times 10^{2}$\\ 
        \hline
        $c\bar{c}$ & [-7, 7] & xFitter & $1.42\times 10^{2}$ & $1.34\times 10^{2}$\\ 
        \hline
        \hline
        $c\bar{c}$ & [-2, 2] & GRV & $1.78\times 10^{2}$ & $1.19\times 10^{2}$\\ 
        \hline
        $c\bar{c}$ & [-2, 2] & JAM21 & $1.39\times 10^{2}$ & $8.36\times 10^{1}$\\ 
        \hline
        $c\bar{c}$ & [-2, 2] & xFitter & $8.59\times 10^{1}$ & $4.91\times 10^{1}$\\ 
        \hline
        \hline
        $c\bar{c}$ & [2, 4.5] & GRV & $6.86\times 10^{1}$ & $6.97\times 10^{1}$\\ 
        \hline
        $c\bar{c}$ & [2, 4.5] & JAM21 & $3.64\times 10^{1}$ & $3.97\times 10^{1}$\\ 
        \hline
        $c\bar{c}$ & [2, 4.5] & xFitter & $2.43\times 10^{1}$ & $3.12\times 10^{1}$\\ 
        \hline
        \hline
        \hline
        $b\bar{b}$ & [-7, 7] & GRV & $2.29\times 10^{0}$ & $4.32\times 10^{0}$\\ 
        \hline
        $b\bar{b}$ & [-7, 7] & JAM21 & $2.11\times 10^{0}$ & $3.47\times 10^{0}$\\ 
        \hline
        $b\bar{b}$ & [-7, 7] & xFitter & $1.54\times 10^{0}$ & $2.50\times 10^{0}$\\ 
        \hline
        \hline
        $b\bar{b}$ & [-2, 2] & GRV & $1.68\times 10^{0}$ & $1.61\times 10^{0}$\\ 
        \hline
        $b\bar{b}$ & [-2, 2] & JAM21 & $1.62\times 10^{0}$ & $1.50\times 10^{0}$\\ 
        \hline
        $b\bar{b}$ & [-2, 2] & xFitter & $1.16\times 10^{0}$ & $1.05\times 10^{0}$\\ 
        \hline
        \hline
        $b\bar{b}$ & [2, 4.5] & GRV & $3.48\times 10^{-1}$ & $9.79\times 10^{-1}$\\ 
        \hline
        $b\bar{b}$ & [2, 4.5] & JAM21 & $2.90\times 10^{-1}$ & $7.02\times 10^{-1}$\\ 
        \hline
        $b\bar{b}$ & [2, 4.5] & xFitter & $2.06\times 10^{-1}$ & $5.41\times 10^{-1}$\\ 
        \hline
        \hline
    \end{tabular}
    \caption{Predictions for the charm and bottom photoproduction cross-section associated with a leading neutron in $pp$ collisions at $\sqrt{s} = 13$ TeV and $pPb$ collisions at $\sqrt{s} = 8.1$ TeV. Results derived assuming different rapidity ranges and using distinct pion PDF parameterizations.}
    \label{tab:total}
\end{table}

In Table \ref{tab:total} we present our results for the charm and bottom photoproduction cross-sections  with a leading neutron in $pp$ collisions at $\sqrt{s} = 13$ TeV and $pPb$ collisions at $\sqrt{s} = 8.1$ TeV, derived assuming different rapidity ranges and using distinct pion PDF parameterizations. In particular, predictions for the rapidity range covered by typical central ($-2.0 \le Y \le 2.0)$ and forward ($2.0 \le Y \le 4.5)$ detectors are presented.   We have that the $pPb$ predictions are $\approx 3$ orders of magnitude larger than for $pp$ collisions. Moreover, the bottom cross-sections are one order of magnitude smaller than for the charm production. Finally, our results indicate that the predictions are sensitive to the pion gluon distribution assumed in the calculations.
{ Such conclusion strongly motivates the improvement of the theoretical description of the process, as well the extension of our analysis for the  production of  identified heavy-flavour-hadrons, as e.g. $D$- and $B$-mesons.} 

\begin{figure}[t]
	\centering
\includegraphics[width=0.47\textwidth]{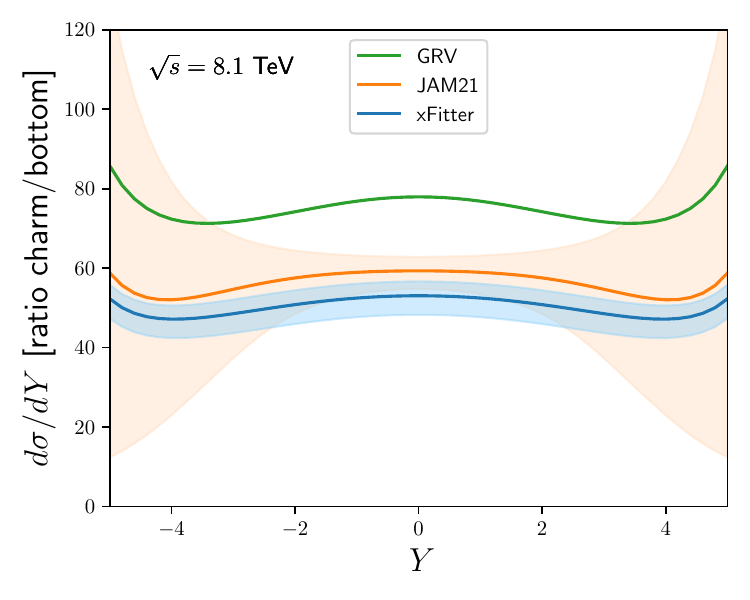}
\includegraphics[width=0.47\textwidth]{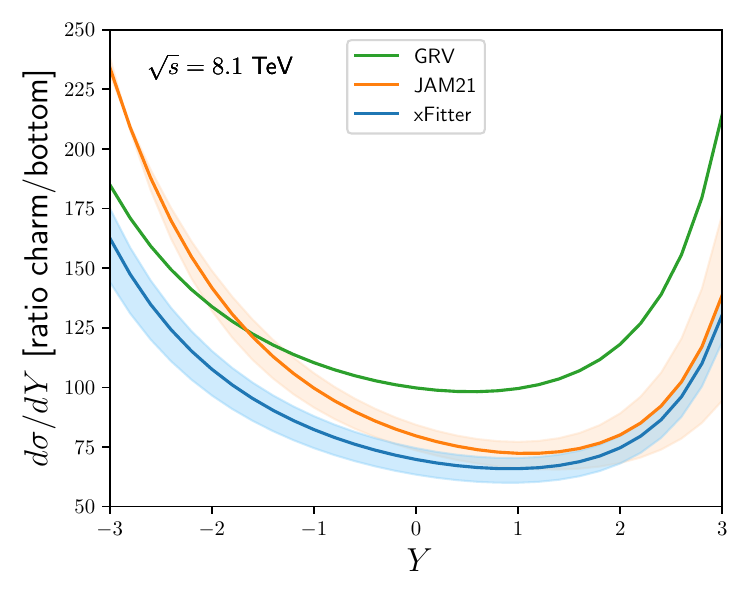}
\caption{Predictions for the ratio between the charm and bottom rapidity distributions in $pp$ (left panel) and $pPb$ (right panel) collisions. Results derived assuming distint pion PDFs. }
\label{fig:ratio}
\end{figure}

Some final comments are in order. The results presented above indicated that a future experimental analysis of the heavy quark photoproduction associated with a leading neutron can be useful to constrain the behavior of the pion gluon distribution at small - $x$. However, a natural question is how reliable are our predictions, since they are dependent on the modeling of the absorptive effects and pion flux as well as of the order of the perturbative calculation  and value assumed for the hard scale. 
As discussed in the previous section, the description of the absorptive effects has been improved in recent years and its magnitude can, in principle, be constrained in a phenomenological way using the HERA, as performed in the current paper. Similarly, the pion flux can also be constrained using the HERA and LHC data for leading neutron processes. On the other hand, the photon-pion cross-section can be evaluated at next-to-leading order, which would reduce the dependence of the predictions on the choices for the factorization and renormalization scales. Such a task is currently being performed. We have verified that the normalization of our predictions is  dependent on the model assumed for the pion flux and on the values for ${\cal{K}}$ and $\mu$, but the difference associated with the distinct pion PDFs is not affected, i.e., our main conclusion remains valid. One way to circumvent this dependence, while the description of main ingredients in the calculation is being improved,  is to consider the ratio between the rapidity distributions for the charm and bottom production. Such a ratio is not dependent on the absorptive effects, and we have verified that the predictions are not affected by the value of $\mu$ and by the order of the perturbative calculation. The corresponding predictions for $pp$ and $pPb$ collisions are presented in Fig.~\ref{fig:ratio}. We have that the ratio is also dependent on the pion PDF and that its analysis in $pPb$ collisions can be very useful to impose additional constrains in the description of the pion structure at high energies.

\section{Summary}
\label{sec:sum}
Despite the theoretical and experimental advances during the last decades, a comprehensive understanding of the partonic structure of the pion remains incomplete. In particular, the behavior of the sea quarks and gluon distributions at small values of the Bjorken $x$ variable is currently poorly understood. In a near future, the  Electron-Ion Collider (EIC) and the proposed Electron-ion collider in China (EicC) are expected to measure the pion structure functions assuming the Sullivan process and tagging the leading neutron, providing additional constrains in these distributions. 
In this paper we have considered an alternative to probe the pion structure at the existing Large Hadron Collider by studying photon-induced interactions, which become dominant in ultraperipheral hadronic collisions. We have proposed the analysis of the heavy quark photoproduction associated with a leading neutron   and carried out an exploratory study of the charm and bottom production
in $pp$ and $pPb$ collisions. The rapidity distributions and cross-sections have been estimated assuming different parameterizations for the pion gluon distribution. We predicted large cross-sections and demonstrated that the rapidity distribution is sensitive to the behavior of the pion gluon distribution at small values of $x$. In addition, we have proposed the analysis of the ratio between the charm and bottom rapidity distributions, which is almost independent of the assumptions for the absorptive effects and the modeling of the pion flux. Our results indicate that the proposed process is a promising way to improve our understanding of the pion structure.  Such results strongly motivate the improvement of the theoretical description of the process as well its implementation in a Monte Carlo event generator. Both subjects will be explored in forthcoming publications.

\begin{acknowledgments}
 V.P.G. and J. T. de Souza were partially supported by CNPq, CAPES (Finance Code 001), FAPERGS and INCT-FNA (Process No. 408419/2024-5).
\end{acknowledgments}

\hspace{1.0cm}

\end{document}